\documentclass[prl,twocolumn, showpacs,preprintnumbers,amsmath,amssymb]{revtex4}

\usepackage{graphicx,color, xfrac, placeins}
\usepackage{multirow}
\usepackage{graphicx}
\usepackage{xcolor}
\usepackage{hyperref}
\usepackage{amssymb}
\usepackage{mathrsfs}
\hypersetup{
    colorlinks,
    linkcolor={blue!},
    citecolor={blue!},
    urlcolor={blue!}
}

\begin{document}

\title{Resonant and inelastic Andreev tunneling\\
observed on a carbon nanotube quantum dot}

\author{J.~Gramich}
\author{A.~Baumgartner}
\email{andreas.baumgartner@unibas.ch}
\author{C.~Sch{\"o}nenberger}
\affiliation{Department of Physics, University of Basel, Klingelbergstrasse 82, CH-4056 Basel, Switzerland}

\begin{abstract}
We report the observation of two fundamental sub-gap transport processes through a quantum dot (QD) with a superconducting contact. The device consists of a carbon nanotube contacted by a Nb superconducting and a normal metal contact. First, we find a single resonance with position, shape and amplitude consistent with the theoretically predicted resonant Andreev tunneling (AT) through a single QD level. Second, we observe a series of discrete replicas of resonant AT at a separation of $\sim145\,\mu$eV, with a gate, bias and temperature dependence characteristic for boson-assisted, inelastic AT, in which energy is exchanged between a bosonic bath and the electrons. The magnetic field dependence of the replica's amplitudes and energies suggest that two different bosons couple to the tunnel process.
\end{abstract}


\maketitle

Electron transport in nanostructures with superconducting (S) and normal metal (N) contacts have attracted considerable attention recently, because of the discovery of new transport processes and exotic quantum mechanical states of matter. While S contacts have been used in bias spectroscopy for many years, transport in nanostructures at energies below the superconductor's energy gap became relevant only recently, for example in the search for solid-state versions of Majorana Fermions \cite{Mourik_Kouwenhoven_Science_2012}, Cooper pair splitting \cite{Hofstetter2009, Herrmann_Kontos_Strunk_PRL104_2010, Hofstetter_Baumgartner_PRL107_2011, Schindele_Baumgartner_Schoenenberger_PRL109_2012}, or the study of Andreev bound states (ABSs) \cite{Pillet_Yeyati_Joyez_NaturePhys6_2010, Dirks_Mason_NaturePhys_2011, Lee_DeFranceschi_NatureNanotech_2013, Schindele_Baumgartner_PRB89_2014}.

The most fundamental low-energy transport process between an N and an S reservoir is Andreev reflection, in which an electron from N can only enter S by forming a Cooper pair with a second electron of opposite spin and momentum (for a standard s-wave superconductor). This process is slightly more complicated if the two reservoirs are connected by a quantum dot (QD), in which the charging energy suppresses a double occupation. The resulting resonant Andreev tunneling (resonant AT) is illustrated in Fig.~1a as an intuitive sequential tunneling process. At zero bias the two electrons forming a Cooper pair tunnel through the QD at the same energy, which leads to a peak in the differential conductance $G$. At a finite bias, resonant AT is allowed only for electrons with energies aligned to the electrochemical potential $\mu_{\rm S}$ of S, i.e., only if the QD resonance ($\mu_{\rm QD}$) is aligned to $\mu_{\rm S}$, see Fig.~1a. Resonant AT can be identified by the distinctive resonance line shape with a sharper decay than when tunneling into a normal reservoir \cite{Beenakker_PRB46_1992, Sun_PRB59_1999, Zhu_PRB64_2001}.

\begin{figure}[b]
	\begin{center}
	\includegraphics[width=0.5\textwidth]{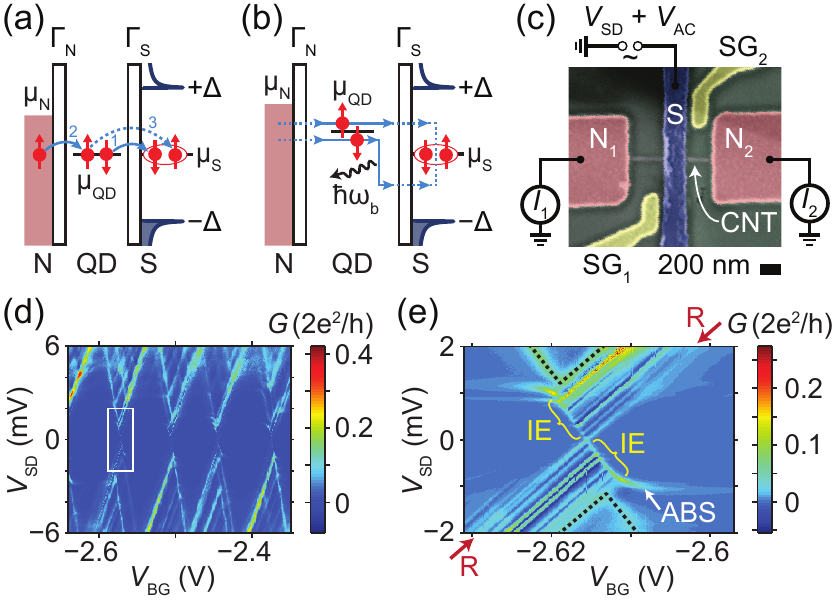}	
\end{center}
	\caption{(Color online) (a) Schematic of resonant AT. Two electrons tunnel through the QD into S for $\mu_{\rm QD}=\mu_{\rm S}$ and $\mu_{\rm N}>\mu_{\rm QD}$. (b) Schematic of inelastic AT. Two electrons tunnel through the QD into S at $\mu_{\rm QD}=\mu_{\rm S}+\frac{n}{2}\hbar\omega_{\rm b}$ for $\mu_{\rm N}>\mu_{\rm QD}$ by emitting multiples of the energy $\hbar\omega_{\rm b}$ to the environment. (c) SEM image of a representative device and schematic of the measurement setup. (d) Differential conductance of QD1 as function of $V_\mathrm{SD}$ and $V_\mathrm{BG}$. (e) Detailed conductance map of the region indicated in (d). The resonant AT line and the inelastic AT are labeled R and IE, respectively. The dotted lines point out the Coulomb blockade diamond edges and ABS an Andreev bound state.}
	\label{fig:SampleOverview}
\end{figure}

An additional transport process is boson-assisted tunneling, in which bosons from the environment are absorbed or emitted in an inelastic tunneling process. This mechanism leads to replicas of the elastic tunnel process at higher energies. For example, in carbon nanotubes (CNTs) with normal metal contacts, phonons can result in additional resonances \cite{Sapmaz_Jarillo-Herrero_van_der_Zant_PRL96_2006} and even in the suppression of the current for strong electron-phonon couplings (Franck-Condon blockade) \cite{Leturcq_Stampfer_Hierold_Ensslin_NaturePhys_2009}. Also the electromagnetic environment can provide energy for photon-assisted processes, resulting in discrete QD resonance side bands \cite{van_der_Wiel_de_Franceschi_Kouwenhoven_Review_2002, Meyer_Kouwenhoven_NanoLetters_2007}. Phonon \cite{Zhang_PRB86_2012} and photon \cite{Sun_PRB59_1999_2} induced discrete replicas of elastic AT were predicted recently for QDs with an S-contact. Such inelastic AT is illustrated in Fig.~1b: a Cooper pair can only be formed by two electrons of opposite energy with respect to the chemical potential of S, $\mu_{\rm S}$. If the QD level is aligned at a positive energy, both electrons traverse the QD at the same energy, and, to form a Cooper pair, one relaxes the energy $\hbar\omega_{\rm b}$ to the environment with a bosonic excitation spectrum. As is evident from Fig.~1c this condition is met at $\mu_{\rm QD}-\mu_{\rm S}=\frac{n}{2}\hbar\omega_{\rm b}\,(n\in\mathbb{N})$, in contrast to twice this value in boson-assisted processes in devices with normal contacts. Neither resonant nor inelastic AT have been observed in QDs before, probably because both require the QD resonance width to be much smaller than the superconducting energy gap, $\Delta$.

Here we use a CNT quantum dot contacted by a superconducting Nb and a normal metal contact to investigate a discrete subgap resonance fully consistent with resonant AT. In addition, the resonance has multiple, very sharp replicas due to inelastic boson-assisted AT. Nb shows a large energy gap in CNT devices \cite{Schindele_Baumgartner_PRB89_2014, Gaass_Strunk_PRB89_2014}, in our devices up to $~1\,$meV, which allows us to perform bias spectroscopy in the regime of the QD life time broadening being considerably smaller than $\Delta$, which is crucial to identify these processes.

Figure 1c shows a false color scanning electron microscopy (SEM) image, including a schematic of the measurement setup. The CNTs were grown by chemical vapor deposition (CVD) on a Si$^{++}$/SiO$_2$ substrate used as a backgate. Subsequently the surface is treated by an rf-induced hydrogen plasma, similar to \cite{Yang_AdvMaterials_2010, Schindele_Baumgartner_PRB89_2014}. In this process about 30\% of the CNTs are etched away, which probably serves as a selection process for defect-free clean CNTs. On a suitable CNT we fabricate a central $\sim200\,$nm wide and $\sim2\,$mm long Ti/Nb ($3\,$nm/$40\,$nm) contact (S) and two normal metal Ti/Au ($5\,$/$65\,$nm) contacts (N) on either side of S at a distance of $\sim 300\,$nm, in a Cooper pair splitter geometry \cite{Schindele_Baumgartner_Schoenenberger_PRL109_2012} with sidegates (SGs) for individual electrical tuning of the two CNT sides. Between the contacts two separate QDs form, but no signals could be found that depend on both QDs \cite{Suppl}. Here we focus solely on experiments on QD1. The experiments were carried out in a dilution refrigerator at a base-temperature of $\sim 110\,$mK.
\begin{figure}[t]
	\begin{center}
	\includegraphics[width=0.5\textwidth]{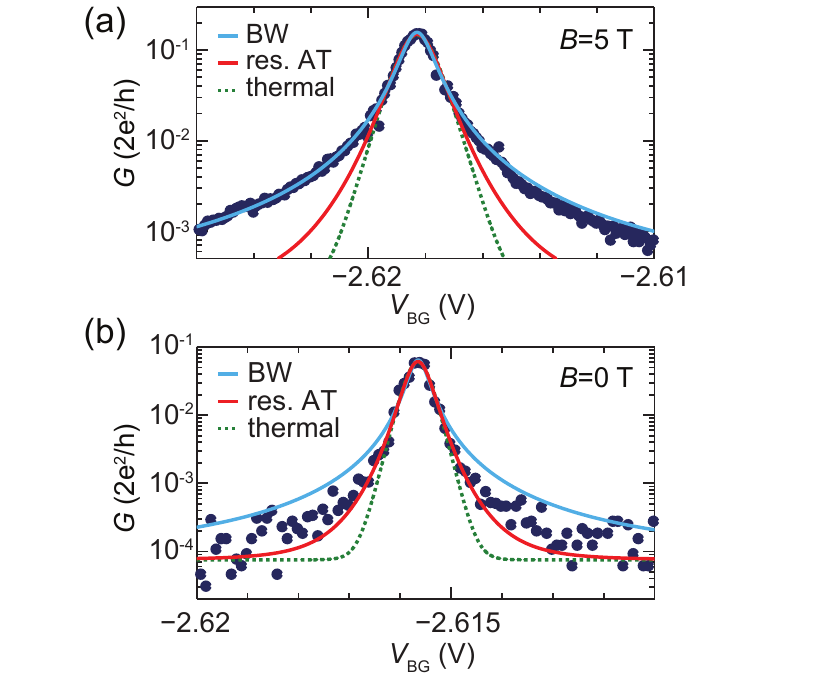}	
\end{center}
	\caption{(Color online) Resonance line shapes: $G$ as function of $V_\mathrm{BG}$ (points) at zero bias (a) in the normal state at $B=5\,$T and (b) at zero field in the superconducting state. The blue, red and dotted green lines are best fits to the data using the expressions for a Breit-Wigner (BW), resonant AT and a thermally broadened QD resonance, respectively.}
	\label{fig:Lineshapes}
\end{figure}
In Fig.~1d the differential conductance $G$ through QD1 is plotted as a function of the backgate voltage, $V_{\rm BG}$, and the bias applied to S, $V_{\rm SD}$. We find very clear Coulomb blockade (CB) diamonds with an addition energy of $\sim 6.6\,$meV, a level spacing of $\sim 1.2\,$meV, and a well-defined superconducting gap of $\sim 1.2\,$meV, which separates the CB diamonds in a characteristic way due to the gap in the single-particle density of states in the S contact \cite{Suppl}. This gap is reduced monotonically with a magnetic field $B$ applied perpendicular to the CNT and the substrate and vanishes at $B\approx 1.5\,$T \cite{Suppl}. Fig.~1e shows $G$ plotted in the region indicated by the rectangle in Fig.~1d. Here we find up to $7$ weak but very sharp parallel resonance lines with the same positive slope as the CB diamonds. Similar features appear in all other CB diamonds \cite{Suppl}, independent of the QD charge state. The average spacing between the lines is $\sim145\pm30\,\mu$eV, with some resonances showing significant deviations from this value. Only one of these parallel lines (R) crosses the entire transport gap, while the other lines (IE) end at a finite sub-gap bias. Pronounced negative differential conductance (NDC) values occur between the conductance maxima. An additional structure arises due to an Andreev bound state (labeled ABS in Fig.~1e), which leads to some amplitude modulation, but will not be discussed in more detail here. Line R we identify with resonant AT. To demonstrate this, we compare the line shape of the zero-bias CB resonance measured in the normal state at $B=5\,$T shown in Fig.~2a to the resonance at zero field, i.e. in the superconducting phase of S, which is plotted in Fig.~2b. For the following fits we added an identical background determined by the data points far off resonance (not shown). In the normal state we expect a Breit-Wigner (BW) line shape due to life-time broadening, which is described by $G(\Delta V_{\rm BG}) =\frac{e^2}{h}\cdot \frac{\Gamma_1\Gamma_2}{\Delta E^2 + \Gamma^2/4}$, with $\Gamma=\Gamma_1 + \Gamma_2$ and the QD level detuning $\Delta E=-e\alpha (\Delta V_{\rm BG}-V_{\rm BG}^{(0)})$, where $\alpha$ is the lever arm of the backgate to the QD and $V_{\rm BG}^{(0)}$ the position of the resonance. This expression fits the observed line shape in Fig.~2a very well (blue curve) for the tunnel coupling parameters $\Gamma_1\approx 9.0\,\mu$eV and $\Gamma_2\approx 96.5\,\mu$eV.
As a comparison, we also plot the best fit using the expression for thermally broadened resonances (dotted green line), $G = \frac{e^2}{h}\frac{1}{4k_{\rm B}T}\frac{\Gamma_1 \Gamma_2}{\Gamma}\cosh^{-2}\left(\frac{\Delta E}{2kT}\right)$, which does not describe the data well. For resonant AT in the limit of non-interacting electrons the expected line shape is $G =\frac{2e^2}{h} \left(\frac{2\Gamma_1\Gamma_2}{4\Delta E^2+\Gamma_1^2+\Gamma_2^2}\right)^2$ \cite{Beenakker_PRB46_1992, Zhu_PRB64_2001}, which also deviates strongly from the data. In contrast, the resonance at zero magnetic field plotted in Fig.~2b is described best by the expression for resonant AT: the measured conductance values decay faster away from the maximum than in the normal state. The extracted values for the tunnel coupling using the BW expression are a factor of $3 - 7$ smaller than in the normal state, while the values extracted from the resonant AT expression, $\Gamma_1\approx 8.4\,\mu$eV and $\Gamma_2\approx 66.5\,\mu$eV, are very similar to the ones in the normal state. We find similar values also in gate sweeps at a small bias \cite{Suppl}.
From the line shapes and the observation that this resonance occurs when the QD level is aligned to the electrochemical potential of S, we conclude that resonance R running through the full energy gap of the superconductor is due to resonant AT.


\begin{figure}[t]
	\begin{center}
	\includegraphics[width=0.5\textwidth]{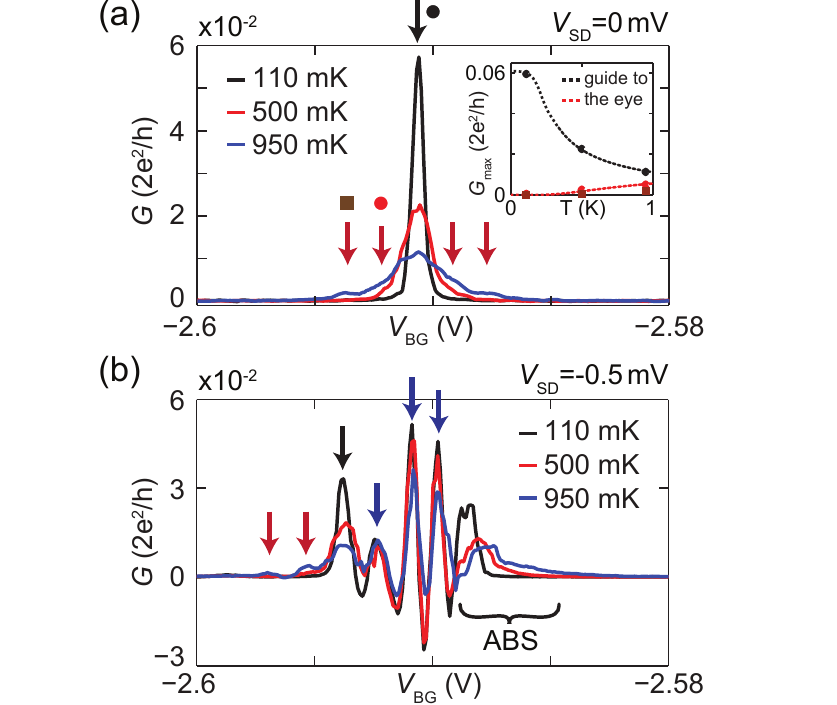}	
\end{center}
	\caption{(Color online) Temperature dependence of resonant and inelastic AT. Differential conductance $G$ as function of $V_\mathrm{BG}$ at (a) $V_\mathrm{SD}=0\,$mV and (b) $V_\mathrm{SD}=-0.5\,$mV for the three indicated temperatures. The black arrow points out the resonant AT line, the red arrows the boson-absorbing and the blue arrows the boson-emitting resonances \cite{Suppl}. The inset in Fig.~(a) shows the peak amplitudes of the features indicated in the main figure.}
	\label{fig:Tdependence}
\end{figure}

The replicas (IE) parallel to the resonant AT line (R) found in Fig.~1e we attribute to the emission and absorption of energy from or into a bosonic reservoir in an inelastic Andreev tunneling process. The bias condition for this process (see Fig. 1b) is $\mu_{\rm S} < \mu_{\rm QD} < \mu_{\rm N}$ \cite{Zhang_PRB86_2012, Suppl}, which results in the lines ending at a finite sub-gap bias, as observed in Fig. 1e. Following the predictions of Ref.~\cite{Zhang_PRB86_2012} for phonons, Fig.~3 shows a series of backgate sweeps obtained at different temperatures. The zero-bias curves in Fig.~3a are symmetric with respect to the central resonant AT peak indicated by the black arrow. The amplitude of this peak is reduced considerably with increasing temperature, while the width increases because of side peaks (red arrows) emerging as weak shoulders between $0.5\,$K and $1\,$K. The amplitudes of the indicated features are plotted in the inset of Fig.~3a for the three temperatures, which shows that resonant AT is reduced with increasing temperature ($\sim 1/k_{\rm B}T$) due to the thermal broadening of the Fermi functions in the normal metal contact. In contrast, the absorption side peaks increase in amplitude at higher temperatures due to the thermal population of the boson states \cite{footnote1}, in good qualitative agreement with the predictions for phonon-assisted AT \cite{Zhang_PRB86_2012, footnote2}.

Backgate sweeps at the bias $V_{\rm SD}=-0.5\,$mV are plotted in Fig.~3b for the same temperatures, which again shows the reduction of the resonant AT amplitude (black arrow) and the onset of boson-assisted inelastic AT at more negative gate voltages (red arrows). At more positive voltages, i.e. $\mu_{\rm S}<\mu_{\rm QD}$, resonances occur with slightly decreasing amplitudes and essentially constant widths (blue arrows) \cite{Suppl}. In addition, we observe pronounced negative differential conductance (NDC) between the peaks. All these findings are in very good qualitative agreement with inelastic AT in which the excess energy of the tunneling electrons is emitted into a bosonic bath \cite{Zhang_PRB86_2012}. In particular, the resonance width in this process is mainly determined by the tunnel couplings, which we do not expect to change greatly with temperature. NDC is also expected, since the resonance condition is determined by the QD level position and not by the Fermi energy of the normal metal lead, which leads to peaks in the current and a peak-dip structure in the differential conductance.


We now investigate the resonant and inelastic AT as a function of an external magnetic field $B$ applied perpendicular to the substrate plane. Figure~4a shows $G$ as a function of $V_{\rm SD}$ and $B$ at the backgate voltage at which the elastic AT resonance (R) crosses $V_{\rm SD}=0$. We find that the energy gap detected by the QD shrinks monotonously up to $\sim0.3\,$T and is then roughly constant up to $\sim1\,$T and diappears around $~1.5\,$T \cite{Suppl}. Here we focus only on the sub-gap features: the resonant AT line R is essentially unaffected at fields below $B\approx 0.7\,$T and splits at higher fields due to Zeeman shifts of the resonances. These do not affect the following analysis since we extract all values directly from Coulomb diamond experiments \cite{Suppl}. We can visualize the transition from resonant AT to a Breit-Wigner characteristics with increasing field by plotting the extracted tunnel parameter, $\Gamma=\Gamma_1+\Gamma_2$, as a function of $B$, which is shown in Fig.~4b. We assume that the tunnel couplings stay roughly constant with increasing $B$ \cite{Suppl}. While up to $B\approx 1\,$T the expression for resonant AT reproduces nicely the large-field values of the BW fit (light red band), it overestimates the coupling by more than a factor of $2$ at high fields. At intermediate fields between $1\,$T and $3\,$T both line shapes deviate considerably from the expected values, which suggests that normal electron tunneling and AT co-exist at fields where the visible transport gap is reduced to zero.

Figure~4c and 4d show the magnetic field dependence of the position (energy) and the peak amplitude of the inelastic AT lines, respectively, extracted from CB spectroscopy \cite{Suppl}. We plot the position of the conductance maxima relative to the resonance R at negative and positive bias up to $1\,$T and label the lowest resonances in energy in Fig.~4c. While the energy of resonance IE1 is essentially constant in this field interval, the positions of IE2, IE3 and IE4 are all reduced and scale similarly with increasing field, but not linear with the energy gap \cite{Suppl}. For the latter resonances the spacing stays roughly the constant. Even more pronounced is the difference in the field dependence of the resonance amplitudes, plotted in Fig.~4d: the amplitude of IE1 is independent of $B$ within experimental error, but the amplitudes of IE2 and IE3 decay continuously on the scale of $1\,$T.

\begin{figure}[t]
	\begin{center}
	\includegraphics[width=0.5\textwidth]{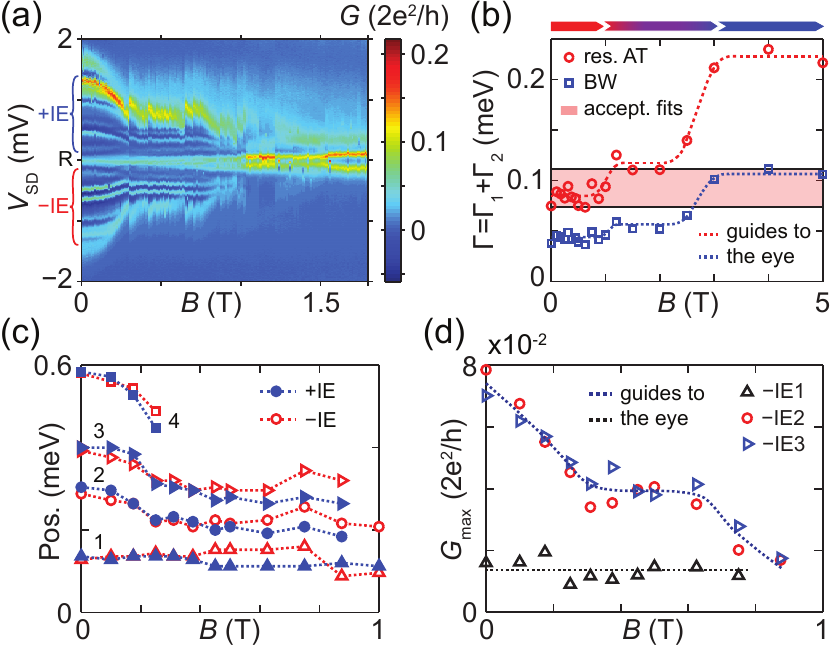}	
\end{center}
	\caption{(Color online) Magnetic field dependence of resonant and inelastic AT. (a) $G$ as a function of the bias $V_{\rm SD}$ and the external magnetic field $B$ at a fixed backgate voltage where the resonance (R) crosses $V_{\rm SD}=0$. (b) Tunnel coupling $\Gamma=\Gamma_1 + \Gamma_2$ of resonance (R) extracted from the resonant AT and the Breit-Wigner expressions as a function of $B$. The fits are obtained from colorscale images always at the same relative position \cite{Suppl} (c) Energy and (d) amplitude of the inelastic AT peaks (IE) as function of $B$. The resonances are labeled in Fig.~c for increasing energy.}
	\label{fig:Bdependence}
\end{figure}

Inelastic AT is mediated by the absorption or emission of bosons. In our system, three types of bosons might be responsible for these sub-gap processes: (I) phonons, e.g. mechanical oscillations in the CNT \cite{Zhang_PRB86_2012}, where the longitudinal acoustic modes in the doubly-clamped CNT, or the CNT squash mode \cite{Dresselhaus_PhysRep_2005} can yield quantized energies of the observed energy scale. (II) plasma modes in the millimeter scale superconducting contact \cite{Camarota_PRL86_2001}. (III) photons, i.e. electromagnetic modes of the resonator formed by the inductance and capacitance of the S-contact, which are damped by ohmic dissipation in normal metals. Phonons naturally account for the temperature dependence of the inelastic AT lines, but it is not straight-forward to explain a magnetic field dependence. Photons would, at least qualitatively, account for the field and temperature dependence, since the kinetic inductance $L_{\rm k}$ of S diverges with decreasing energy gap, which results in a reduced resonance frequency $\propto (L_{\rm k}C)^{-\sfrac{1}{2}}$ ($C$ is the capacitance to the backgate) \cite{Watanabe_JJAP33_1994}. The fact that we find qualitatively different magnetic field characteristics for the lowest energy inelastic AT peak compared to the resonances at higher energies might indicate that two different bosonic baths are coupled to our QD.


In summary, the large superconducting transport gap found in our Nb-contacted CNT QDs and the sharp QD resonances allow us to identify resonant (elastic) and inelastic Andreev tunneling in a QD-superconductor structure. In addition, we also demonstrate the impact of the energy gap on the CB diamond structure. The temperature dependence of the inelastic replicas of resonant AT is consistent with bosonic excitations that open additional transport channels. However, from our experiments the nature of the bosons is difficult to assess. The magnetic field dependence might even hint at the possibility of two different bosonic systems coupling to the QD. Our experiments demonstrate that, in contrast to normal metal systems, such excitations can become the dominant transport mechanisms in S-QD systems, while a smearing of the discrete resonances is a possible origin of the heavily discussed 'soft gaps' \cite{Pekola_PRL105_2010} in large-gap superconductor nanostructures. In addition, the coupling to bosonic reservoirs is also expected to result in replicas of other sub-gap features, for example ABSs \cite{Baranski_Domanski_arXiv_1503.07119_2015}. Similar devices with engineered bosonic environments could shed more light on the nature of the observed processes and lead to hybrid quantum systems with a controlled coupling to the superconductor. For example, novel techniques allow the fabrication of suspended CNTs with superconducting and other contact materials \cite{Gramich_Baumgartner_PSS_2015}, which should give access to well-controlled CNT phonons \cite{Benyamini_Ilani_NaturePhys_2014}. Using the same technique, a radio-frequency cavity can be coupled to the CNT QD, which results in a well-controlled electromagnetic environment with discrete modes \cite{Ranjan_Jung_Schoenenberger_NatureComm_2015}.

We thank C. Stampfer and S. Csonka for helpful discussions and J. Schindele and M. Weiss for their support in the laboratory. We also gratefully acknowledge the financial support by the EU FP7 project SE$^2$ND, the EU ERC project QUEST, the Swiss NCCR Quantum and the Swiss SNF.

\bibliographystyle{naturemag}


\clearpage
\onecolumngrid
\section{Supplementary material}

\renewcommand\thefigure{S\arabic{figure}}  
 
\section{\label{section:TheoryNQDS}Quasi-particle tunneling in an N-QD-S system} 

\begin{figure}[b]
	\begin{center}
	\includegraphics[width=\textwidth]{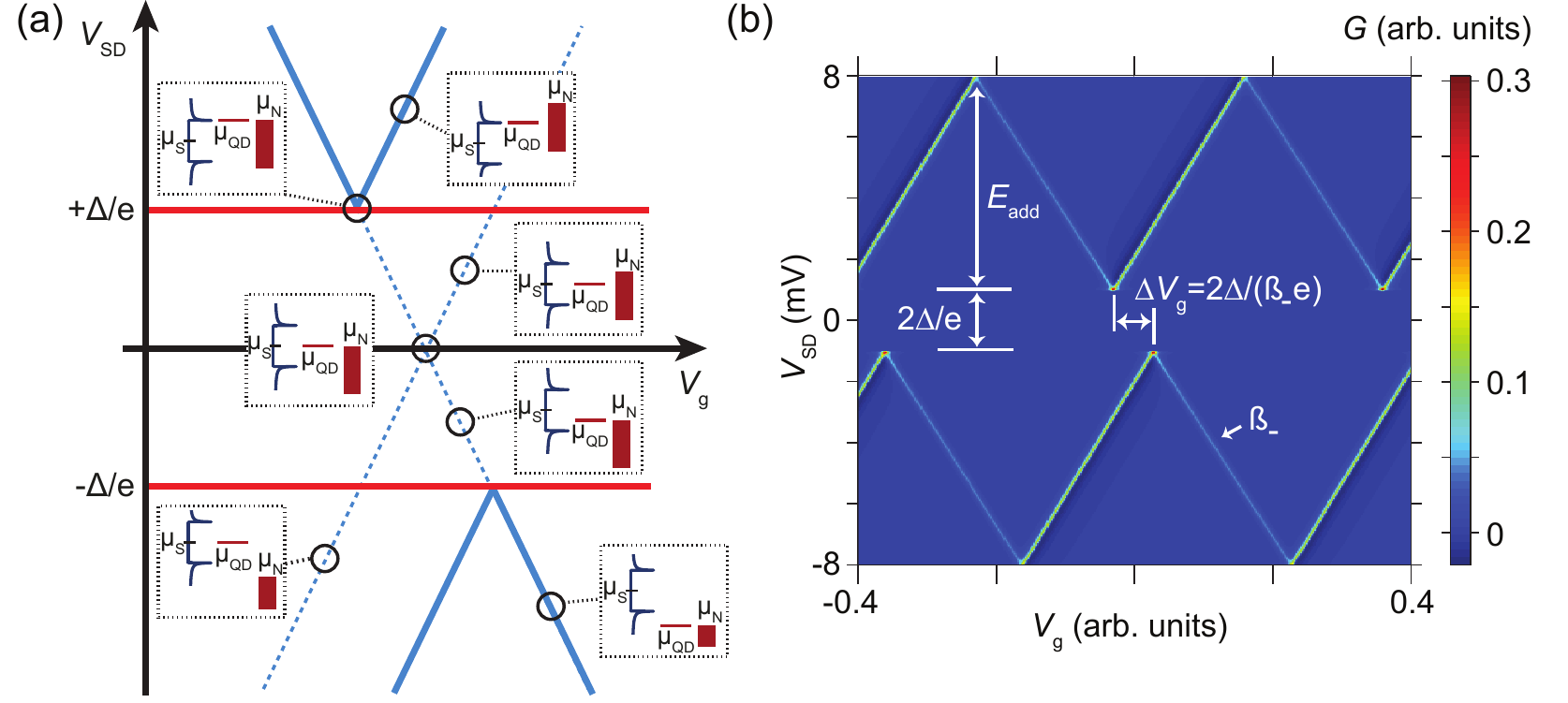}	
\end{center}
	\caption{(a) Relative alignment of the electrochemical potentials of the superconductor $\mu_\mathrm{S}$, normal contact $\mu_\mathrm{N}$ and quantum dot $\mu_\mathrm{QD}$ for the bias applied to S. (b) $G$ vs $V_\mathrm{SD}$ and $V_\mathrm{g}$ obtained from Eq. (\ref{eq:Toymodel}) with $\Gamma_1=10\,\mu$eV, $\Gamma_2=100\,\mu$eV, $E_\mathrm{add}=7\,$meV and $\Delta=1\,$meV at $T=100\,$mK.}
	\label{fig:TheoryNQDS}
\end{figure}

We briefly discuss quasi-particle tunneling in a S-QD-N structure. The corresponding experimental results can be found in Fig. 1d and 1e of the main text and in Fig. \ref{fig:CharacterizationQD1}. Figure \ref{fig:TheoryNQDS} sketches the differential conductance for a quantum dot (QD) connected to a normal metal (N) and a superconducting contact (S) with the bias applied to S (see Fig. 1a main text). Figure \ref{fig:TheoryNQDS}a shows the relative positions of the electrochemical potentials $\mu_\mathrm{N}$, $\mu_\mathrm{S}=-eV_\mathrm{SD}+\mu_\mathrm{N}$ and $\mu_\mathrm{QD}$ for selected positions of the expected charge stability diagram. The tips of the Coulomb blockade (CB) diamonds mark the onset of quasiparticle tunneling, separated in bias by $2\Delta/e$. The tips are shifted in gate voltage with respect to each other due to the capactive coupling of the leads to the QD by $\Delta V_\mathrm{g}=2\Delta/(e\beta_-)$, with $\beta_-=-C_\mathrm{g}/(C_\mathrm{tot}- C_\mathrm{S})$ the (negative) slope of the $\mu_\mathrm{N}=\mu_\mathrm{QD}$ CB resonance. $C_\mathrm{S}$, $C_\mathrm{N}$, $C_\mathrm{g}$, $C_\mathrm{tot}=C_\mathrm{S}+C_\mathrm{N}+C_\mathrm{g}$ are the S, N, gate and total capacitances, respectively. If one considers only single electron transport through the system and neglects superconducting correlations and charge dynamics on the QD, the qualitative charge stability diagram  can be understood by a simple resonant tunneling picture. The current is given by\cite{Datta:1995}
\begin{equation}
I \propto \int{dE \mathscr{D}_\mathrm{N}(E)\mathscr{D}_\mathrm{S}(E+eV_\mathrm{SD})\cdot T_\mathrm{QD}(E)\cdot [f_\mathrm{N}(E)-f_\mathrm{S}(E+eV_\mathrm{SD})]}.
\label{eq:Toymodel}
\end{equation}
The resulting differential conductance $G=\mathrm{d}I/\mathrm{d}V_\mathrm{SD}$ is plotted in Fig. \ref{fig:TheoryNQDS}b which reproduces the qualitative arguments made in Fig. \ref{fig:TheoryNQDS}a. Here we chose the density of states (DOS) $\mathscr{D}_\mathrm{N}(E)=1$, a BCS-type DOS for the superconductor $\mathscr{D}_S(E)=|E|/(\sqrt{E^2-\Delta^2})$ for $|E|\geq \Delta$ and $\mathscr{D}_S(E)=0$ for $|E|<\Delta$, and a Breit-Wigner transmission function for the QD of the form\cite{Ihn:2010} $T(E)=(\Gamma_1\Gamma_2)/(\Delta E^2 + (\Gamma_1+\Gamma_2)^2/4)$ with $\Delta E = E-\mu_\mathrm{QD} \pm n\cdot E_\mathrm{add}\,(n \in \mathbb{N})$, which also accounts for gating of the single QD level by $\mu_\mathrm{QD}=\mu_\mathrm{QD}^0-e\frac{C_\mathrm{S}}{C_\mathrm{tot}}V_\mathrm{SD}-e\frac{C_\mathrm{g}}{C_\mathrm{tot}}V_\mathrm{g}$ and the electron filling with the addition energy $E_\mathrm{add}$.

\FloatBarrier

\section{\label{section:PAqpTunneling}Boson-assisted quasi-particle and Andreev tunneling in an N-QD-S system}

\begin{figure}[h]
	\begin{center}
	\includegraphics[width=0.9\textwidth]{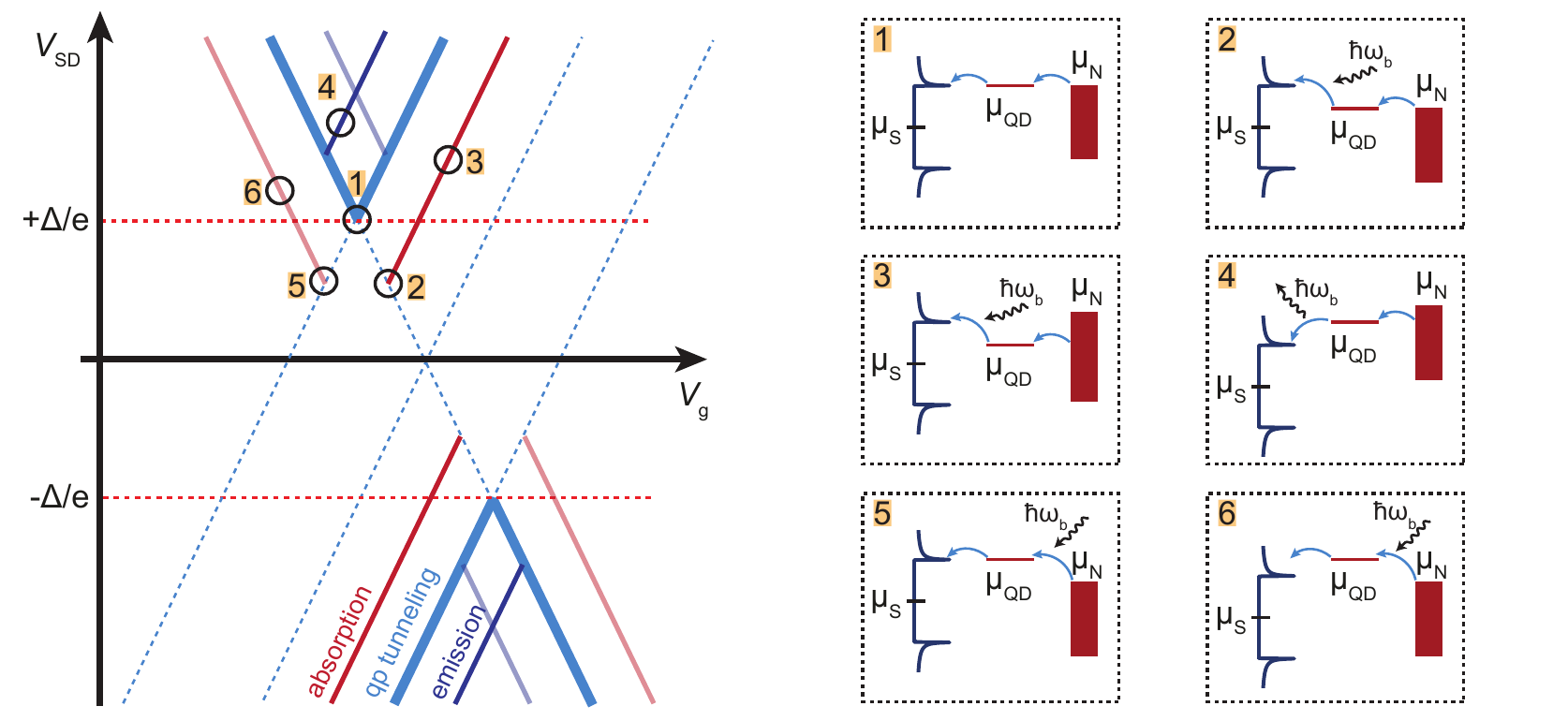}	
\end{center}
	\caption{Expected gate and bias dependence for boson-assisted quasi-particle tunneling in an N-QD-S system considering only single-boson processes. Blue lines denote emission, red lines absorption processes. Replicas of the gap edge can be either due to boson processes on the S side (positive slope) or on the N side (more transparent lines, negative slope). The energy diagrams to the right depict schematically the processes at specific gate and bias voltages.}
	\label{fig:PAqpTunneling}
\end{figure}

\begin{figure}[t]
	\begin{center}
	\includegraphics[width=0.9\textwidth]{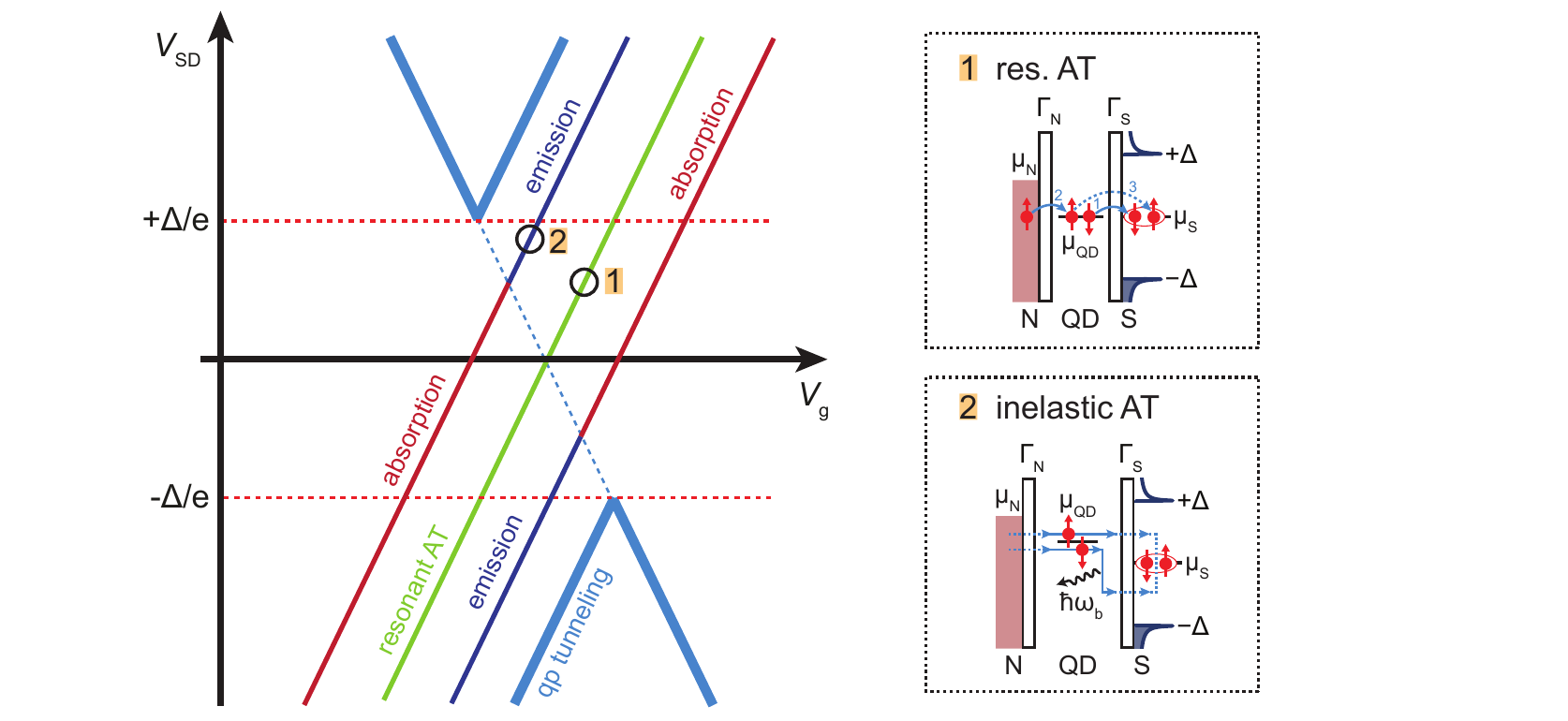}	
\end{center}
	\caption{Expected gate and bias dependence of resonant (green line) and boson-assisted Andreev tunneling. Replicas of resonant Andreev tunneling can occur  due to boson emission (blue lines) and boson absorption (red lines). }
	\label{fig:ResATDiagram}
\end{figure}

The subgap resonances parallel to the Coulomb diamond edges observed in our experiments (features IE and R in Fig.~1e of the main text) are due to resonant (R) Andreev tunneling (AT), with replicas due to boson-assisted, inelastic (IE) AT. Here we discuss in addition inelastic quasi-particle (qp) tunneling. \cite{Deblock:2003} In this process a quasi-particle can be transferred from one lead to the other by the absorption or emission of a boson. To compare boson-assisted qp and Andreev tunneling, we briefly sketch the expected signatures of these processes in a charge stability diagram, see Fig.~\ref{fig:PAqpTunneling} for boson-assisted qp tunneling and Fig.~\ref{fig:ResATDiagram} for resonant and inelastic Andreev tunneling. The energy diagrams to the right describe the processes at some characteristic gate voltages.

Inelastic AT results in replicas parallel to the resonant Andreev tunneling line (R) with a spacing given by $\hbar\omega_\mathrm{b}/2$, while in boson-assisted qp tunneling the Coulomb blockade diamond edges are replicated with a spacing determined by $\hbar\omega_\mathrm{b}$. A crucial difference between the two transport mechanisms is the arrangement of the boson-absorbing (red lines in diagrams) and boson-emitting resonances (blue lines). For boson-assisted qp tunneling shown in Fig.~\ref{fig:PAqpTunneling}, the absorption replicas are inside the Coulomb blockaded regions and the emitting lines are above the Coulomb diamonds, similar to the structures observed for a QD contacted by normal metal contacts.\cite{Leturcq:2009}
In contrast, the gate dependence of the Andreev processes is shown in Fig.~\ref{fig:ResATDiagram}. Resonant AT results in a resonance crossing the entire energy gap between the Coulomb diamond edges with a positive slope (green line). The boson-assisted replicas run parallel to this line, but in contrast to qp tunneling, the boson-emission and absorption lines merge where they meet the line connecting the Coulomb diamond edges with negative slope (see also Ref.~\citenum{Zhang:2012}).

The nature of the observed resonances can now be determined based on the temperature dependence. With increasing temperature, the amplitude of an absorption line should increase due to the increased boson population, while the one of an emission line should decrease. In our experiments (see Fig. 3 main text and Fig. \ref{fig:Tdependence}), we observe a decrease of conductance for the IE lines with increasing temperature and the appearance of absorption lines connecting to the emission lines, consistent with boson-assisted AT in the schematic of Fig. \ref{fig:ResATDiagram} and calculated in Ref.~\citenum{Zhang:2012}. We find no structures in the experiments related to the ones described in Fig.~\ref{fig:PAqpTunneling}. Furthermore, for qp tunneling one could also expect broader subgap resonances and similar processes at the N-QD interface with different rates, generating another set of replicas sketched in Fig. \ref{fig:PAqpTunneling} as weaker lines.

\section{\label{section:Setup}Experimental setup and shielding}

The experiments were performed in an Oxford Kelvinox dilution refrigerator with an electron temperature of $T\sim 100-150\,$mK. To minimize thermal radiation from warmer stages of the setup and the effects due to quasi-particles, \cite{Barends_APL99_2011} we filter the electrical lines using pi-filters at room temperature and a tape-worm filter between the 1K pot and the mixing chamber. On the coldfinger directly above the sample holder additional pi-filters are employed to dampen also the 1K radiation. The sample is shielded from the helium bath by a Faraday box and two additional copper shields. Since the electronic temperature is considerably lower than the critical temperature of the Nb strip (see section \ref{section:NbStrip}) and the observed energy gap (see main text), the generation of quasi-particles is strongly suppressed. Furthermore, three independent cool-downs of the device in two different measurement setups allow us to exclude any spurious effects from the electronic setup (not shown). 

\section{\label{section:CharacterizationSampleTB0}Sample characteristics}

\subsection{\label{subsection:CharacterizationQD1}Characterization of QD 1}

\begin{figure}[h]
	\begin{center}
	\includegraphics[width=8cm]{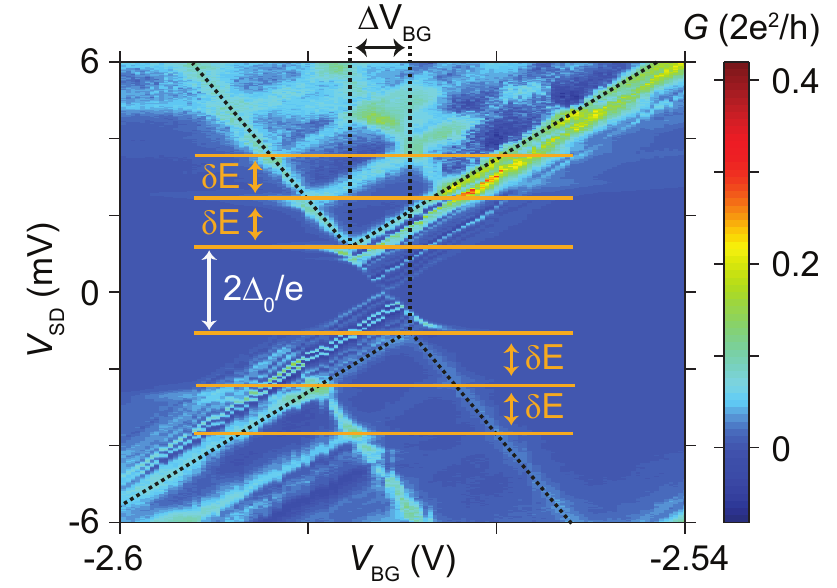}	
\end{center}
	\caption{Differential conductance $G$ of QD1 as function of the bias $V_\mathrm{SD}$ and the global backgate voltage $V_\mathrm{BG}$ at $T=110\,$mK and $B=0\,$T. The orange lines are used to read out the level spacing $\delta E$ from the visible excited states. The white arrow indicates the visible superconducting transport gap $\Delta_\mathrm{0}$. Black dashed lines mark the shifted CB diamonds.}
	\label{fig:CharacterizationQD1}
\end{figure}

Using individual charge stability diagrams as in Fig. 1d of the main text and Fig. \ref{fig:CharacterizationQD1}, we extract a leverarm of the global backgate of $\alpha\approx 90\,$meV/V and an addition energy of $E_\mathrm{add}\approx 6.6\,$meV. The shift of the CB diamonds due to the superconductor dicussed in section \ref{section:TheoryNQDS} and marked in Fig. \ref{fig:CharacterizationQD1} allows us to determine the superconducting transport gap at $B=0$, $\Delta_\mathrm{0}\approx 1.2\,$meV. In Fig. \ref{fig:CharacterizationQD1}, we find conductance resonances running in parallel to the edge of the diamonds for  $|eV_\mathrm{SD}|>\Delta_\mathrm{0}$. These are due to (electronic) excited states of the QD, which exhibit a Zeeman splitting in an external magnetic field\cite{Escott:2010} (see section \ref{section:Bevolution}). From their roughly equidistant spacing in energy (marked in the figure by orange horizontal lines), we find a level spacing of $\delta E \approx 1.2\,$meV, much larger than the spacing of the boson-replicas, $\delta E \gg \hbar\omega_\mathrm{b}$. The characteristic energies in our system can thus be summarized as $kT\ll\Gamma < \hbar\omega_\mathrm{b} \ll \delta E \approx \Delta_\mathrm{0} \ll E_\mathrm{add}$.

\FloatBarrier

\subsection{No Cooper pair splitting or QD hybridization}

\begin{figure}[h]
	\begin{center}
	\includegraphics[width=0.87\textwidth]{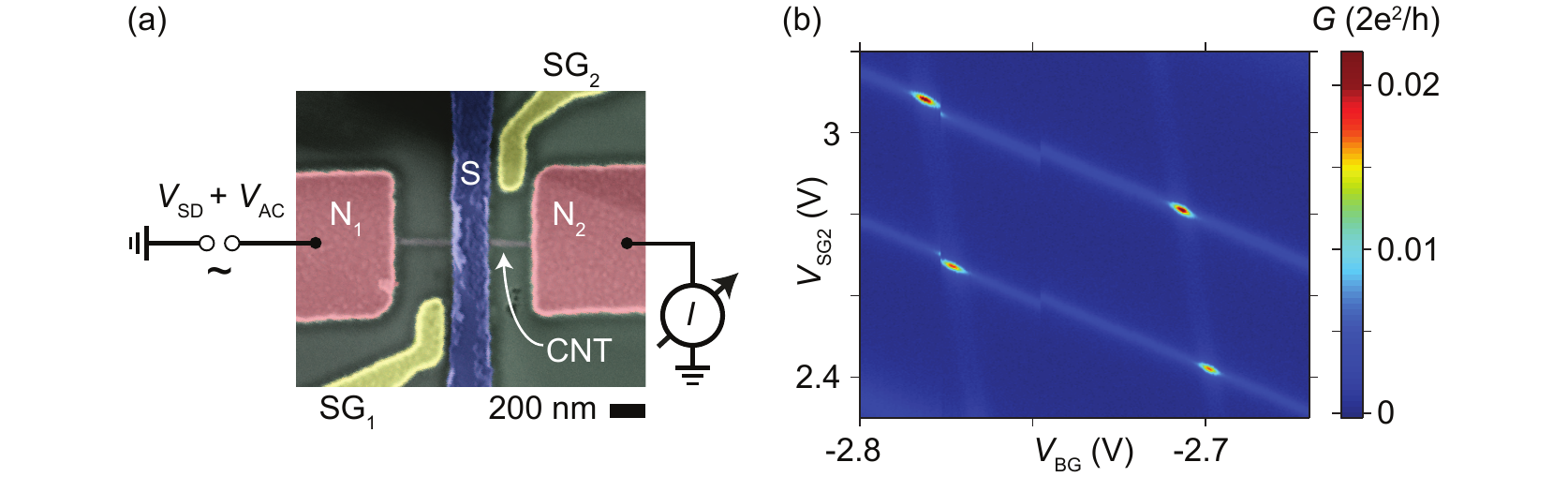}	
\end{center}
	\caption{(a) Experimental setup to measure the two QDs in series. The superconductor is floating (not connected to ground). (b) Serial double-QD conductance map of sidegate 2 voltage vs backgate voltage at $V_\mathrm{SD}=0\,$mV, $B=5\,$T and $T=110\,$mK.}
	\label{fig:NoNonlocal}
\end{figure}

To exclude any effects due to the second QD of the device we measured $G$ vs. sidegate and backgate voltage with the two QDs in series, see Fig. \ref{fig:NoNonlocal}a. The resulting double-QD charge stability diagram in Fig. \ref{fig:NoNonlocal}b confirms that the two QDs are well decoupled by the superconducting contact in between. In particular, we do not observe the typical honeycomb pattern or anti-crossings, but solely an increased conductance at the resonance crossing points. This suggests that the inter-dot coupling is much smaller than the individual QD life times, $\Gamma_{12}\ll \Gamma_\mathrm{QD1},\Gamma_\mathrm{QD2}$, and that also capacitive cross-talk is negligible. In experiments with the two QDs in parallel, we do not find any conductance features that depend on both QDs, excluding effects due to Cooper pair splitting.

\subsection{\label{section:SubgapBT0}Additional analysis of the subgap features}
\begin{figure}[t]
	\begin{center}
	\includegraphics[width=0.86\textwidth]{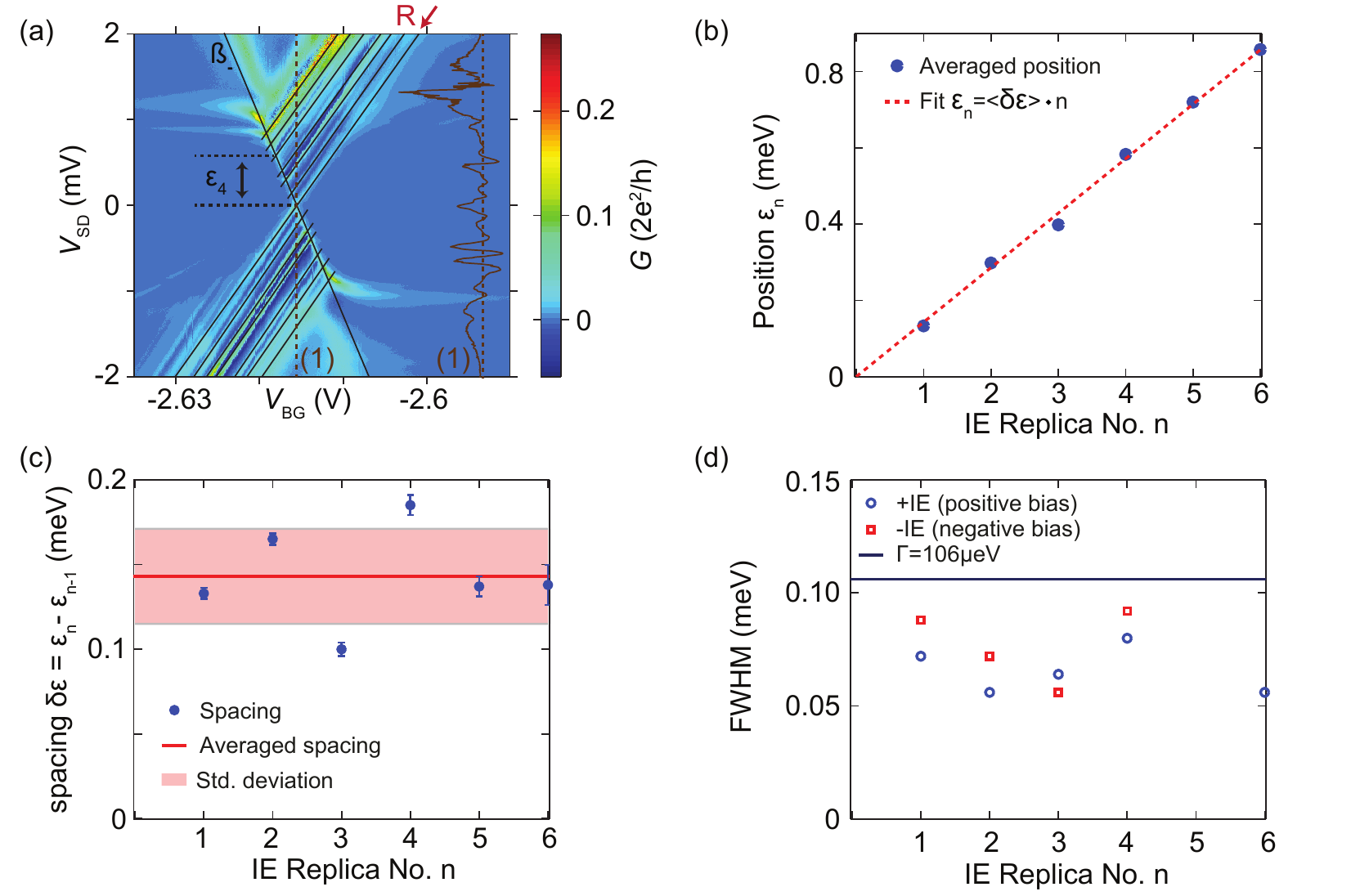}	
\end{center}
	\caption{(a) Read-out of the energy $\epsilon_n$ (and spacing) of the IE subgap features in a $G$ vs backgate and bias voltage map. The crossing point between the negative slope $\beta_-$ of the CB diamond and a black line along the conductance maximum of IE determines the position in energy of the feature relative to the resonant Andreev line (R). The IE4 replica ($n=4$) at positive bias is pointed out as an example. The brown curve shows exemplary a vertical cross-section at the backgate voltage indicated by a dashed line. (b) Averaged position $\epsilon_n$ vs. the IE replica number $n$, data averaged from 2 datasets for positive and negative bias. The red dashed line shows a linear fit $\epsilon_n=<\delta\epsilon>\cdot n$, giving an averaged IE replica spacing of $<\delta\epsilon>=143\,\mu$eV. (c) Spacing $\delta\epsilon=\epsilon_n-\epsilon_{n-1}$ as function of IE replica state number $n$, data averaged over 2 datasets for positive and negative bias. The red line and area mark the averaged spacing $<\delta\epsilon>=143\,\mu$eV and standard deviation determined in (c). (d) FWHM of the IE features as a function of $n$, for positive (blue circles) and negative bias (red squares). The line marks the value for $\Gamma=106\,\mu$eV determined in the main text.}
	\label{fig:SubgapBT0}
\end{figure}
\begin{figure}[t]
	\begin{center}
	\includegraphics[width=0.86\textwidth]{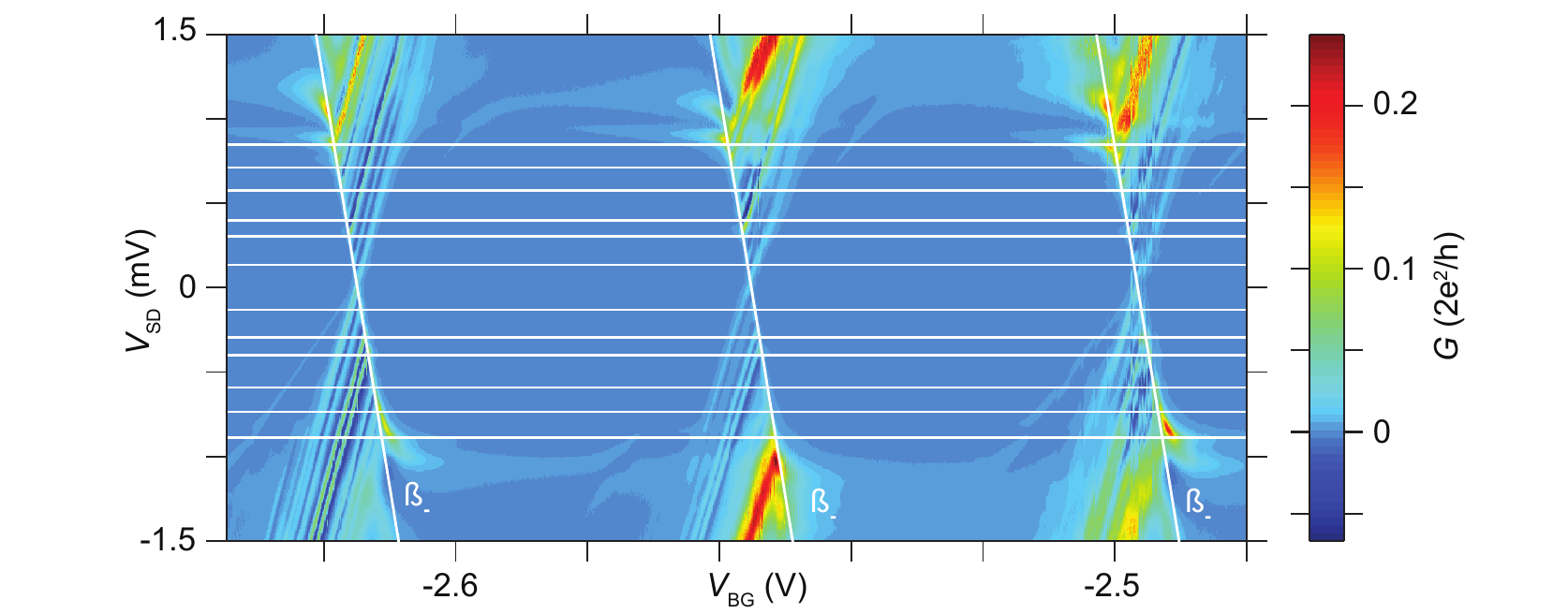}	
\end{center}
	\caption{$G$ as function of $V_\mathrm{BG}$ and $V_\mathrm{SD}$ over 3 Coulomb blockade resonances. The horizontal white lines mark the position $\epsilon_n$ of the IE replica states read out in Fig. \ref{fig:SubgapBT0}, the line with slope $\beta_-$ is drawn for better orientation. The positions of the IE lines are independent of the charge state of the QD within experimental error.}
	\label{fig:IndepElectronState}
\end{figure}
Figure \ref{fig:SubgapBT0} demonstrates the analysis of the subgap features and inelastic replicas (IE) of the resonant Andreev line (R) at $B=0\,$T and $T=110\,$mK. Figure \ref{fig:SubgapBT0}a explicitly shows how we determine the position in energy $\epsilon_n$ of the IE replica lines relative to the resonant line R from conductance maps. The vertical cross-section through the middle of the features is used for the analysis of the conductance maxima $G_\mathrm{max}$ and the full width half maximum (FWHM) of the IE replica lines. This cross-section also demonstrates the pronounced negative differential conductance and Fano-like linehape of the IE features, which agrees with the predictions in Ref.~\citenum{Zhang:2012}. From the analysis of the position and spacing in Fig. \ref{fig:SubgapBT0}b,c we obtain an averaged spacing (over all $n$) between the IE lines of $<\delta \epsilon > = \hbar\omega_\mathrm{b}/2=(143\pm 28)\,\mathrm{\mu eV}$, corresponding to an average boson energy of $\hbar\omega_\mathrm{b}=286\,\mathrm{\mu eV}$.

Even if the IE replica energies scale roughly linear with $n$, some replicas significantly deviate from the average spacing (Fig. \ref{fig:SubgapBT0}c), possibly due to a different origin of the resonances, as discussed in the main text. Here, one also has to bear in mind that the position of $G_\mathrm{max}$ might not completely agree with the position of the current maximum predicted for boson-assisted tunneling,\cite{Zhang:2012} maybe partially explaining the deviations due to the overlap and slightly different widths of neighboring IE replica lines. To obtain an estimate for the width of the IE replica lines, we neglect the influence of negative differential conductance and determine the FWHM from  $G_\mathrm{max}$ to $G=0$, plotted in Fig. \ref{fig:SubgapBT0}d as function of the IE number. We note that the FWHM stays roughly constant and smaller than the determined $\Gamma$ from the fits, i.e. also $\mathrm{FWHM}< \sqrt{2}\Gamma$ as predicted by theory.\cite{Zhang:2012} We obtain an average FWHM of $\sim 70\,\mathrm{\mu eV}$ for the IE lines. In summary, we find average values for the boson frequency of $f_\mathrm{b}\approx 69\,$GHz, a $\mathrm{FWHM}\approx 17\,$GHz and a resonance Q factor of $Q\sim f_\mathrm{b}/\mathrm{FWHM}\approx 4$. Similar to Ref. \citenum{Leturcq:2009}, one can study the maximum conductance of the IE replicas $G_\mathrm{max}$ as a function of the replica line number (number of emitted bosons). We observe a varying $G_\mathrm{max}$ (not shown), with the maximum for the negative bias e.g. at the $n=2$ and $n=3$ lines (visible in Fig. \ref{fig:SubgapBT0}a), reminiscent of a Franck-Condon blockade scenario.\cite{Leturcq:2009}
Figure \ref{fig:IndepElectronState} demonstrates that neither the position nor the spacing of the IE replicas depends on the electronic charge state of the QD. The relative intensities of the replica lines are also similar, even if the total conductance of the replicas change for different CB resonances. 
\section{\label{section:Fitting}Line shape fits at finite bias}
\begin{figure}[h]
	\begin{center}
	\includegraphics[width=8.4cm]{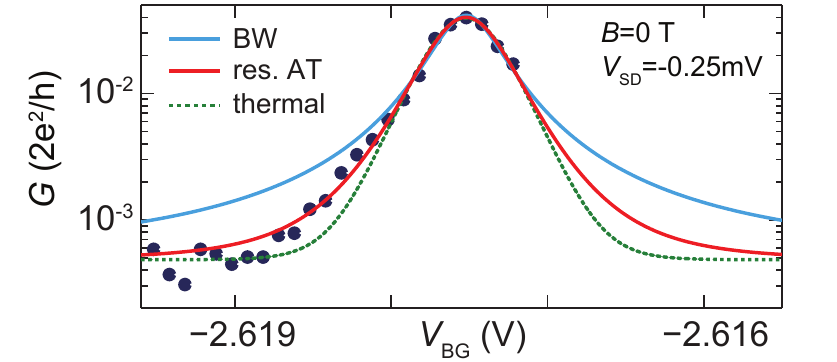}	
\end{center}
	\caption[]{Fits of the resonant Andreev line R similar to the main text at $B=0\,$T and $T=110\,$mK. The data (blue points) is a backgate sweep at $V_\mathrm{SD}=-0.25\,$mV. The data points to the right of the resonance are ignored (not shown) because of the inelastic AT features. The blue, red and dashed green lines are best fits according to the Breit-Wigner, resonant AT and thermally broadenend CB line shape, respectively (see main text).}
	\label{fig:FittingRATfiniteBias}
\end{figure}

To confirm that our fits in Fig. 2b of the main text at $V_\mathrm{SD}=0$ are independent of an overlapping Andreev bound state (ABS), we fit the expressions for a Breit-Wigner, the resonant AT and of a thermally broadened CB resonance to the conductance data at a small bias $|V_\mathrm{SD}|\ll\Delta_0/e$. Figure \ref{fig:FittingRATfiniteBias} shows such fits for a cross-section at $V_\mathrm{SD}=-0.25\,$mV. The resonant Andreev tunneling line shape (red line) agrees very well with the data. From this fit, we extract $\Gamma_1\approx 6.9\,\mathrm{\mu eV}$, $\Gamma_2\approx 68.3\,\mathrm{\mu eV}$ and $\Gamma=\Gamma_1+\Gamma_2\approx 75.2\,\mathrm{\mu eV}$, in very good agreement with the values obtained in the main text for $V_\mathrm{SD}=0\,$mV. Due to the (weak) observed Andreev bound state (see main text), we tentatively attribute the larger tunnel coupling to the superconducting contact, i.e. $\Gamma_{\rm S}=\Gamma_2$. To control our fits in the normal state at larger magnetic fields, we also performed fits at $B=0\,$T and finite bias $|V_\mathrm{SD}|>\Delta_0/e$ of the quasiparticle tunneling lines of the CB diamonds (not shown). Again the Breit-Wigner line shape agrees best with the data and we obtain very similar $\Gamma$-values compared to the fit at $B=5\,$T shown in the main text.
\section{\label{section:Tevolution}Evolution with temperature}

\begin{figure}[h]
	\begin{center}
	\includegraphics[width=0.95\textwidth]{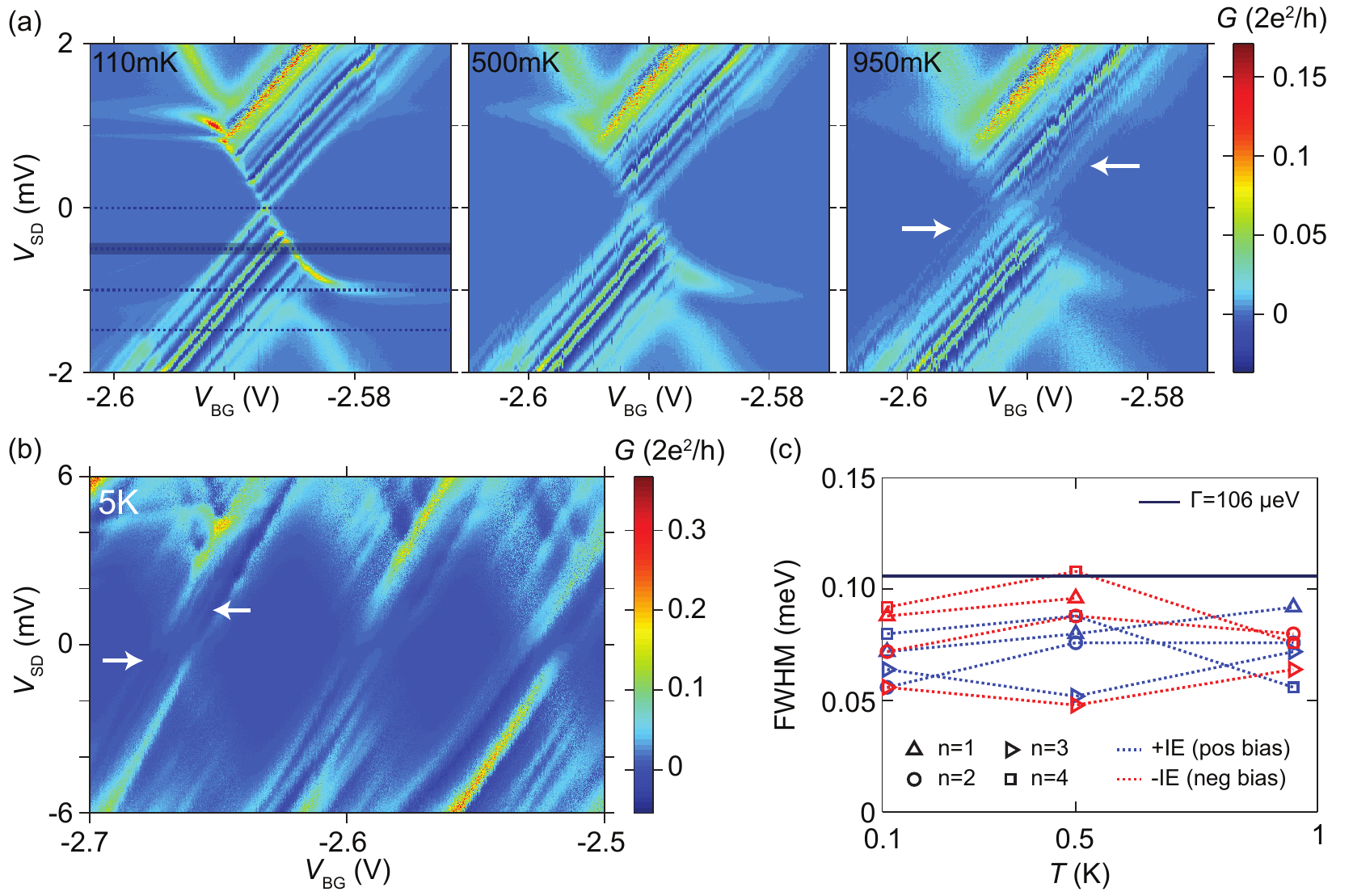}	
\end{center}
	\caption[]{(a) $G$ vs $V_\mathrm{BG}$ and $V_\mathrm{SD}$ at $T=110\,$mK, $T=500\,$mK and $T=950\,$mK. The dark dashed lines mark the position of the studied horizontal cross-sections at fixed $V_\mathrm{SD}$, shown exemplary in the $T=110\,$mK map. The shaded area at $V_\mathrm{SD}=-0.5\,$mV  visualizes the data used for averaging. White arrows in the $T=950\,$mK map point at boson-sidebands due to boson-absorption. (b) Similar conductance map as in (a) at $T\approx 5\,$K. Even at this high temperature similar lines and subgap features are still visible, marked by white arrows. (c) Extracted FWHM vs temperature for the $n=1...4$ IE replicas for positive and negative bias. The FWHM is constant for this temperature range and slightly smaller than $\Gamma=106\,\mathrm{\mu eV}$ extracted for the QD life time broadening (from the main text, dark blue line).}
	\label{fig:Tdependence}
\end{figure}
Figure \ref{fig:Tdependence}a shows $G$ as a function of $V_\mathrm{SD}$ and $V_\mathrm{BG}$ for different temperatures $T=110\,$mK, $T=500\,$mK and $T=950\,$mK. At higher temperatures, $kT \sim \hbar\omega_\mathrm{b}/2$, the conductance maximum of the resonant Andreev line and the emitted-boson replicas start decreasing and boson-absorbed sidebands appear. This characteristics is already evident from the raw data (white arrows in Fig. \ref{fig:Tdependence}a). We study this behavior also in cross-sections at fixed $V_\mathrm{SD}$ values, indicated in Fig. \ref{fig:Tdependence}a. To average out noise occuring at higher temperatures, 12 cross-sections around the desired $V_\mathrm{SD}$ value were averaged, including the compensation of the horizontal shifts due to gating.
Other cross-sections (not shown) show the same features as the $V_\mathrm{SD}=-0.5\,$mV cross-section discussed in the main text. From a similar analysis as in section \ref{section:SubgapBT0} and Fig. \ref{fig:SubgapBT0} we can conclude that both the position and the spacing of the IE replicas stays constant within the studied temperature range. Fig. \ref{fig:Tdependence}c shows the extracted FWHM of the IE replica lines as a function of temperature. The FWHM of the resonant line R (not shown) and the replicas stay roughly constant and $\mathrm{FWHM}\lesssim\Gamma$, in agreement with the prediction\cite{Zhang:2012} $\mathrm{FWHM}< \sqrt{2}\Gamma$ independent of $T$, further supporting our interpretation. Fig. \ref{fig:Tdependence}b shows a conductance map recorded at $T\approx 5\,\mathrm{K} < T_\mathrm{C1}=7.7\,$K (cf. section \ref{section:NbStrip}). Even if most features and the superconducting gap are already smeared with temperature, one still observes pronounced lines and features in the subgap region running parallel to the positive slope of the CB diamonds, marked by white arrows in the figure.

\section{\label{section:Bevolution}Evolution with magnetic field}

\begin{figure}[t]
	\begin{center}
	\includegraphics[width=0.95\textwidth]{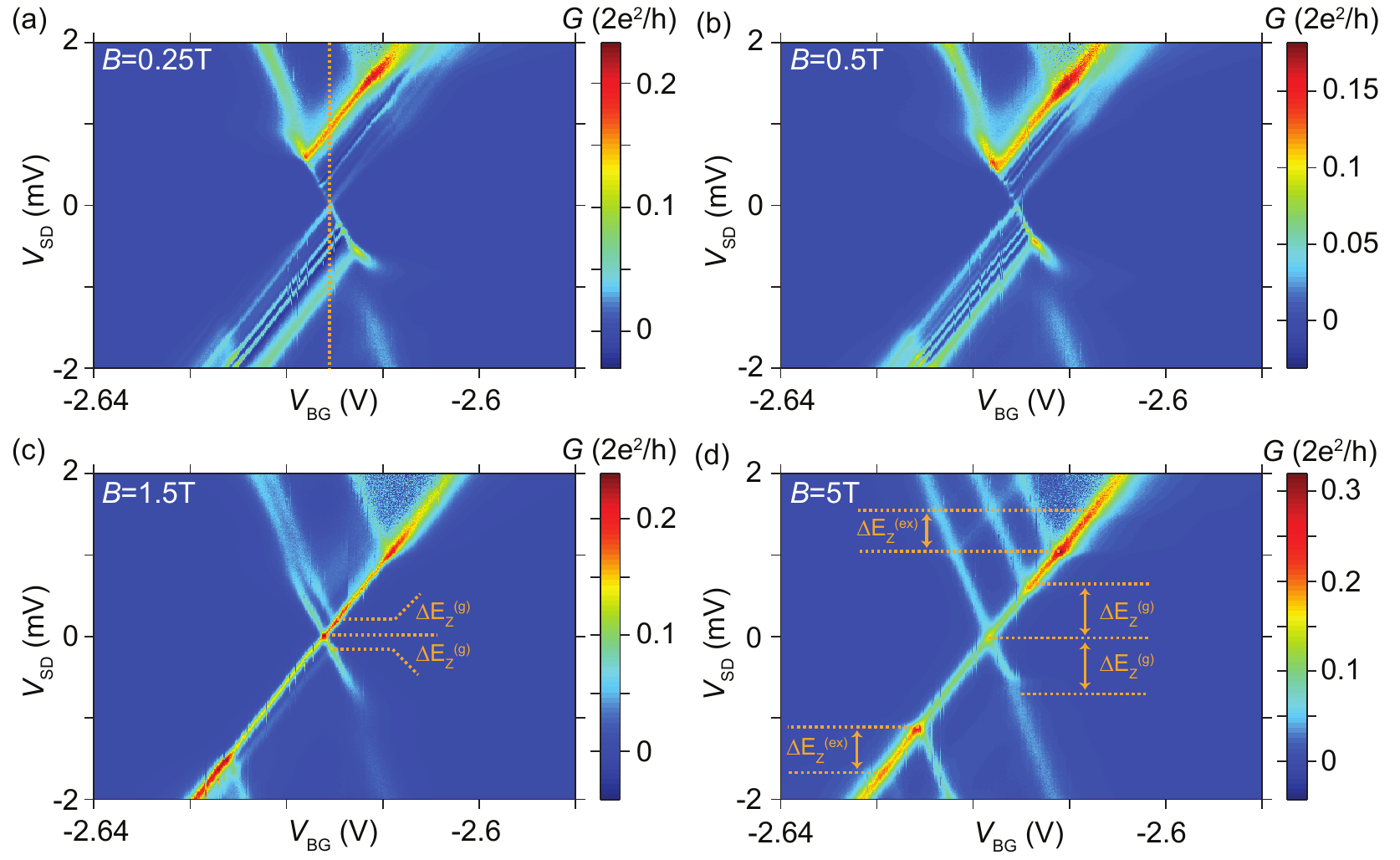}	
\end{center}
	\caption{Example data for the $B$-field analysis: $G$ vs $V_\mathrm{BG}$ and $V_\mathrm{SD}$ colormaps for (a) $B=0.25\,$T, (b) $B=0.5\,$T, (c) $B=1.5\,$T and (d) $B=5\,$T. The magnetic field is applied perpendicular to the sample plane and CNT. The vertical dashed line in (a) indicates the position of the measurement in Fig. 4a of the main text. In (c) and (d), the marked $\Delta E_\mathrm{Z}^\mathrm{(g)}$ and $\Delta E_\mathrm{Z}^\mathrm{(ex)}$ correspond to the Zeeman splitting of the ground and orbital excited states for positive and negative bias, respectively.}
	\label{fig:BRawdata}
\end{figure}

Figure \ref{fig:BRawdata} shows $G$ vs. $V_\mathrm{BG}$ and $V_\mathrm{SD}$ for a series of external magnetic fields applied perpendicular to the sample plane.  From such conductance maps we extract for each field in the main text the following quantities: (i) the transport gap $\Delta$ from the tips of the shifted diamonds as discussed in section \ref{section:TheoryNQDS}, (ii) the Zeeman splitting of the ground and orbital excited states of the QD, indicated in Fig. \ref{fig:BRawdata}d, (iii) the $B$-field dependence of resonant Andreev tunneling from horizontal cross-sections at $V_\mathrm{SD}=0\,$mV, (iv) the position, spacing, conductance maximum $G_\mathrm{max}$ and FWHM of IE replica lines from an analysis similar to section \ref{section:SubgapBT0}. We now discuss some additional results in more detail. 

\begin{figure}[b]
	\begin{center}
	\includegraphics[width=0.96\textwidth]{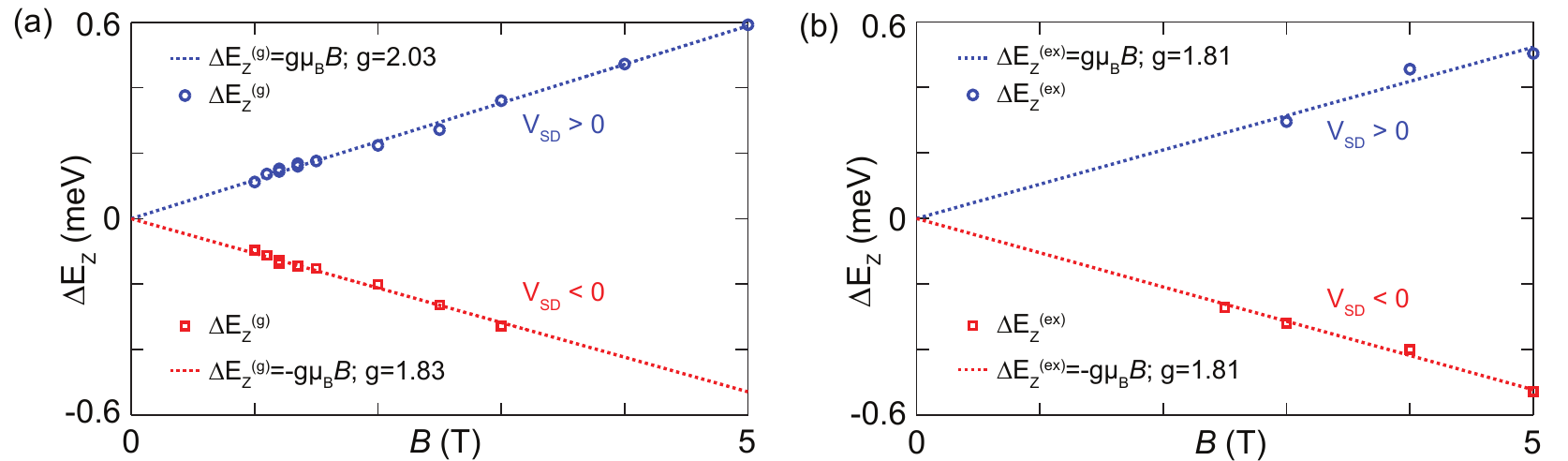}	
\end{center}
	\caption{Zeeman splitting $\Delta E_\mathrm{Z}^\mathrm{(g)}$ (a) and $\Delta E_\mathrm{Z}^\mathrm{(ex)}$ (b) of the ground and first orbital excited state for positive and negative bias, respectively, read-out from the B-field colormaps as demonstrated in Fig. \ref{fig:BRawdata}. We obtain a g-factor of roughly $g\approx 1.9$ from the fits $\Delta E_\mathrm{Z}=g\mu_BB$.}
	\label{fig:Zeeman}
\end{figure}

\begin{figure}[t]
	\begin{center}
	\includegraphics[width=\textwidth]{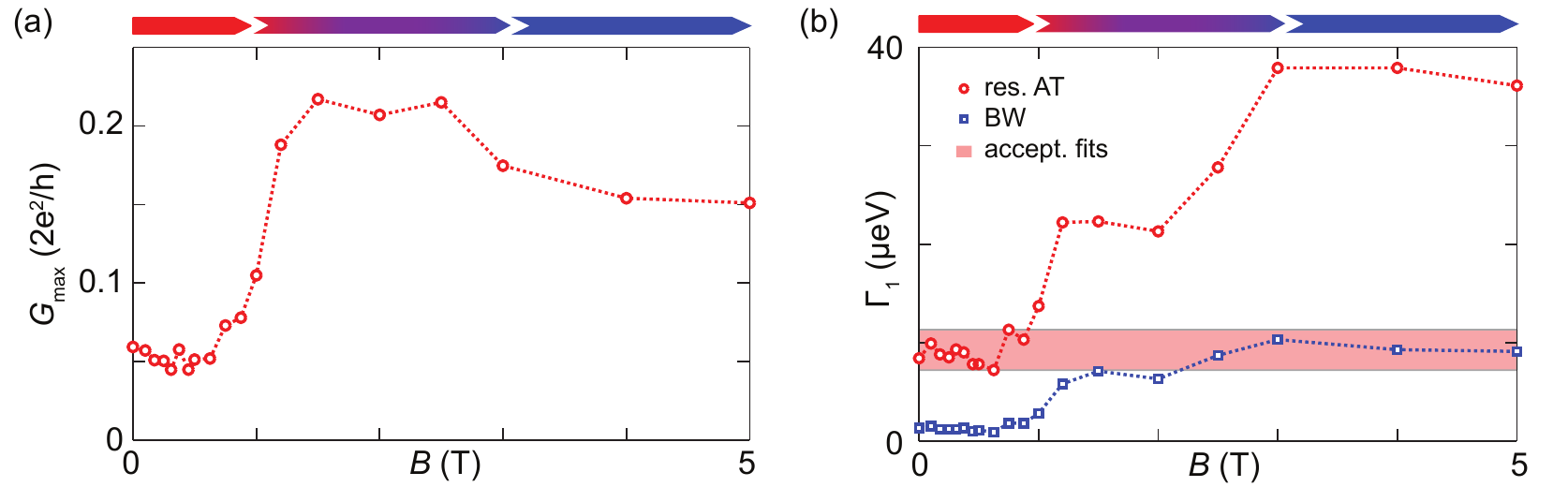}	
\end{center}
	\caption{$B$ dependence of resonant AT. We study horizontal cross-sections at $V_\mathrm{SD}=0\,$mV as function of $B$, extracted from individual colormaps. (a) conductance maximum $G_\mathrm{max}$ of the resonant line (R) vs $B$. (b) $\Gamma_1$ extracted from fits to the expressions for resonant AT (red circles) and the Breit-Wigner line shape (blue squares). The shaded red area indicates the $\Gamma_1$ values for fits in good agreement with the data.}
	\label{fig:RATvsB}
\end{figure}

Figure \ref{fig:Zeeman}a and b show the analysis of the Zeeman splitting $\Delta E_\mathrm{Z}=g\mu_BB$ of the ground and first orbital excited state of QD1. We obtain a g-factor of $g\approx 1.9$. 
Figure \ref{fig:RATvsB} shows the magnetic field dependence of the resonant Andreev tunneling peak R, for (a) the conductance maximum $G_\mathrm{max}$ and (b) the extracted $\Gamma_1$ from individual fits to the resonant AT and the Breit-Wigner (BW) line shape (see main text). $\Gamma_1$ is the most sensitive parameter in these fits. In both plots, we observe a clear transition (marked on top of the plots with a colorscale) with contributions from resonant AT only (red, up to $B\approx 1\,$T) to an intermediate region where both processes coexist ($B\approx 1-3\,$T) to contributions from normal electron tunneling alone (blue, $B\approx 3.5-5\,$T). This is immediately evident when one looks at the range of $\Gamma_1$ for acceptable deviations from the high-field BW values in Fig. \ref{fig:RATvsB}b. The values for small $B$ extracted from resonant AT and  for large $B$ from normal tunneling agree very well. Similarly, one observes an increase in the conductance from $G_\mathrm{max}\sim 0.06\times 2e^2/h$ to $G_\mathrm{max}\sim 0.22\times 2e^2/h$ in the intermediate region where both processes coexist, followed by a decrease of conductance to $G_\mathrm{max}\sim 0.16\times 2e^2/h$ at high fields.
\begin{figure}[b]
	\begin{center}
	\includegraphics[width=\textwidth]{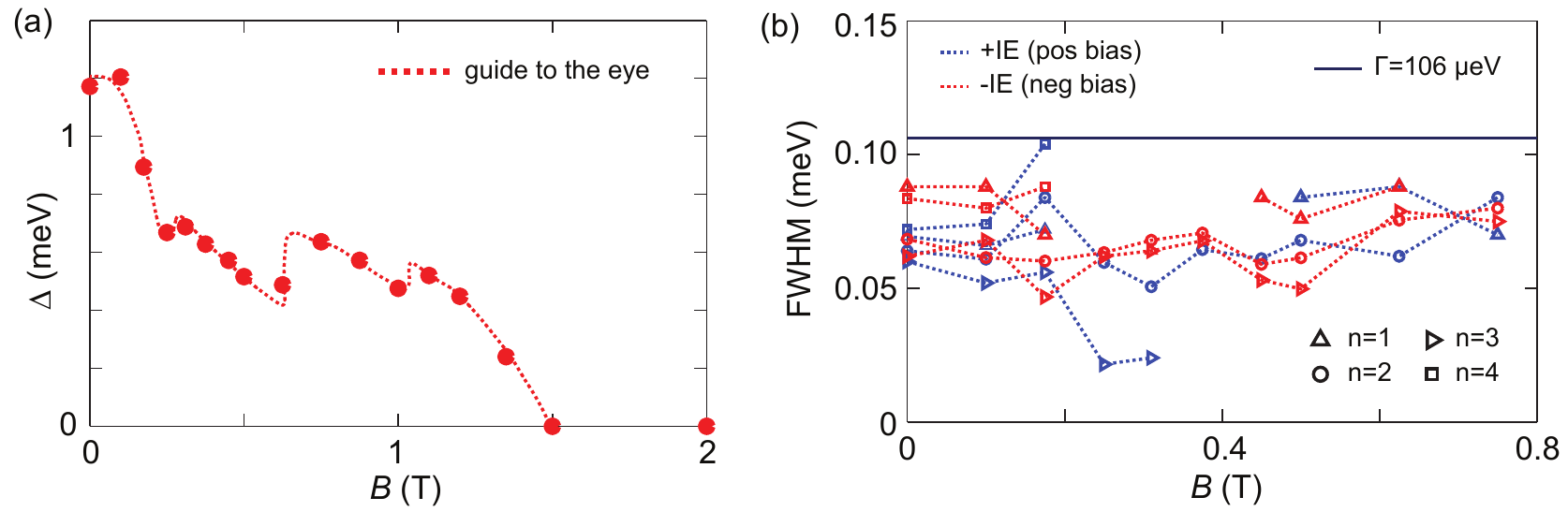}	
\end{center}
	\caption{(a) Transport gap $\Delta$ vs $B$, extracted from individual colormaps similar to Fig. \ref{fig:BRawdata}. The red dashed line is a guide to the eye. (b) Extracted FWHM vs $B$ for the $n=1...4$ IE replicas for positive and negative bias. The FWHM stays constant as function of $B$ and smaller than $\Gamma=106\,\mathrm{\mu eV}$ (from the main text, dark blue line).}
	\label{fig:GapFWHM}
\end{figure}
From the conductance maps, we can also extract the visible transport gap $\Delta$ as function of $B$, shown in Fig. \ref{fig:GapFWHM}a. We find that $\Delta$ is suppressed at $B\sim 1.5\,$T. This field corresponds to the transition region from resonant AT to BW. The detailed characteristics of $\Delta$ seems at first somewhat surprising. $\Delta$ is reduced fast up to $B\sim 0.3\,$T, then stays roughly constant up to $B\sim 1.2\,$T, before it monotonuously tends to zero at $B\sim 1.5\,$T. In the region between $0.3$ and $1.2\,$T, we observe non-monotonuous, systematic fluctuations around a constant value. The positions of this opening-closing behavior of $\Delta$ are quite reproducible in several repeated measurements and also evident in the $V_\mathrm{SD}$ vs $B$ map shown in Fig. 4a of the main text. In addition we observe a hysteresis in $\Delta$ between up- and downsweeps of $B$ (not shown). We ascribe this behavior to vortices in the superconducting type (II) phase of the narrow Nb strip. Indeed, a first critical field $B_\mathrm{C1} \sim 0.3\,$T, marking the initial decay of $\Delta$ and the onset of the hysteresis and `jumps', would be in reasonable agreement with literature values for thin Nb films.\cite{LBdatabase:1993} For larger magnetic fields, vortices could enter the region of the narrow Nb strip close to the CNT. We speculate that, depending on the position and distance of these vortices relative to the `sensing' CNT QD region the visible gap $\Delta$ could be larger or smaller, thus explaining the sudden jumps in $\Delta$ with a vortex rearrangement. 
As an extension to the analysis of the position, amplitude and spacing of the replica lines in a magnetic field in the main text, Fig. \ref{fig:GapFWHM}b shows the extracted FWHM of the IE replica lines as function of $B$. The FWHM stays constant as function of $B$ and $\mathrm{FWHM}<\Gamma$ holds.

\section{\label{section:2ndSample}A second sample}

\begin{figure}[h]
	\begin{center}
	\includegraphics[width=\textwidth]{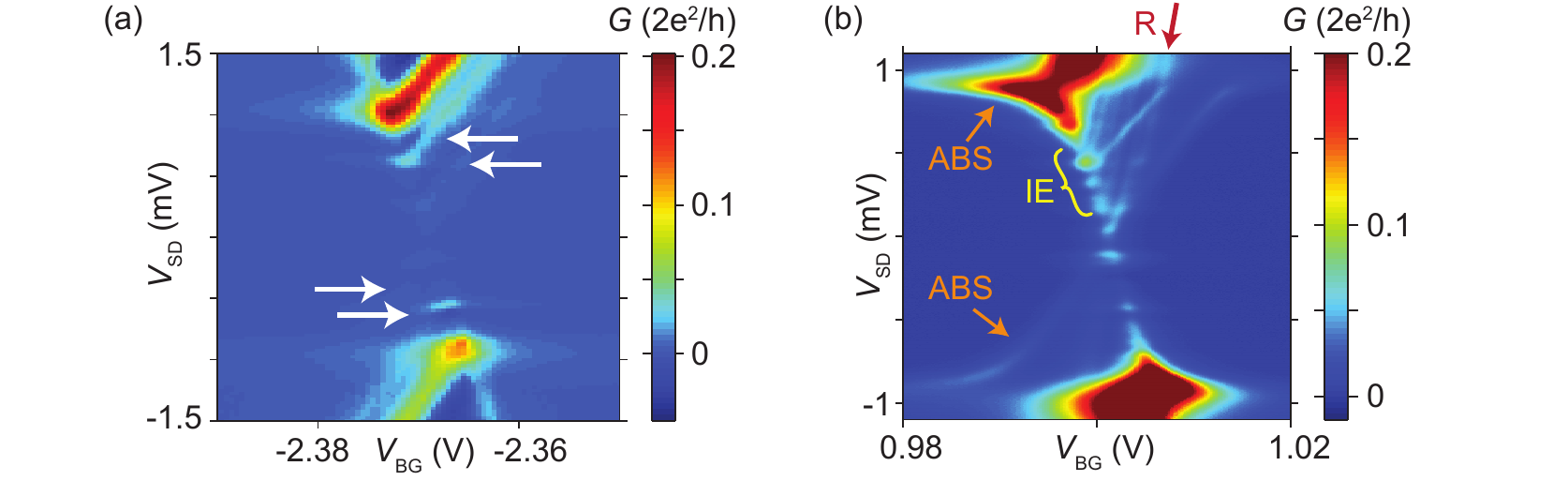}	
\end{center}
	\caption{$G$ vs the backgate voltage and bias $V_\mathrm{SD}$ for (a) QD2 of the same device (b) an independent sample with the same geometry, recorded at $T=60\,$mK. In (a), weak conductance lines running parallel to the CB diamond are labeled with white arrows. In (b), an Andreev bound state (ABS, orange arrows) is more pronounced, but a set of lines running parallel to one edge of the diamond are still visible. The resonant Andreev line R and inelastic replicas IE can be clearly identified here.}
	\label{fig:2ndsample}
\end{figure}

Figure \ref{fig:2ndsample} shows supporting data with very similar findings for QD2 of the same device and for a second sample fabricated in the same geometry. In both cases, one clearly sees straight lines running parallel to the edge of the CB diamond as discussed in detail for the first sample.

\section{\label{section:NbStrip}Characterization of the superconducting niobium strip}

\begin{figure}[h]
	\begin{center}
	\includegraphics[width=\textwidth]{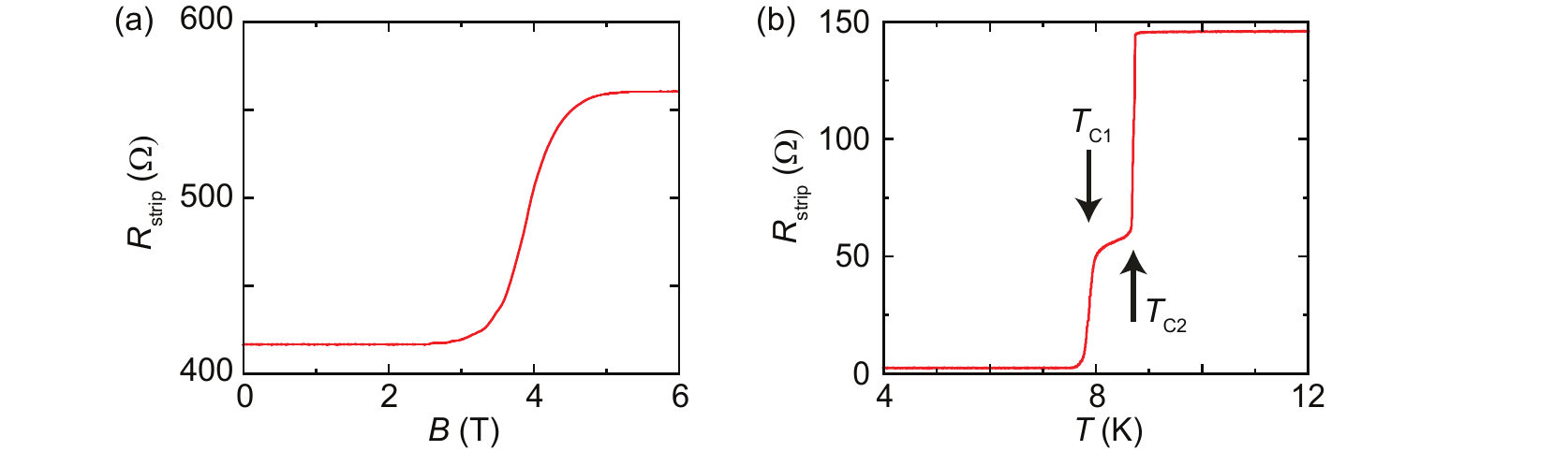}	
\end{center}
	\caption{Resistance of the Nb strip measured in a two terminal configuration as a function of (a) the magnetic field at $T=110\,$mK base temperature, and (b) the temperature in a variable temperature insert (VTI). The two arrows mark two distinct superconducting transition temperatures $T_\mathrm{C1}\sim 7.7\,$K and $T_\mathrm{C2} \sim 8.7\,$K.}
	\label{fig:NbStrip}
\end{figure}

Figure \ref{fig:NbStrip} shows the resistance $R_{\rm strip}$ of the Nb strip acting as S-contact in our devices as function of (a) the external magnetic field, and (b) the temperature, measured in a two terminal geometry with a cryostat line resistance of (a) $415\,\mathrm{\Omega}$ and (b) $2.5\,\mathrm{\Omega}$. The magnetic field dependence in (a) shows a broad transition to the normal state at $\sim 3-4.5\,$T, which coincides with the transition to the pure Breit-Wigner line shape discussed in the main text (cf. section \ref{section:Bevolution}). In (b) we find two distinct superconducting transition temperatures, most likely due to different widths in the design of the superconductor strip. We ascribe the lower $T_\mathrm{C1}\sim 7.7\,$K to the narrow part of the Nb strip in direct contact to the CNT. In the experiments presented in the main text the sample temperature is well below $T_\mathrm{C1}$, so that we can neglect a temperature dependence of the superconducting energy gap.

\FloatBarrier

\bibliographystyle{naturemag}

\begin{thebibliography}{1}
\bibitem{footnote1}
\bibinfo{note}{The conductance maximum is expected to scale with temperature as $\sim 1/k_{\rm B}T\left[ {\rm e}^{\hbar\omega_{\rm b}/k_{\rm B}T}-1\right]^{-1}$}
\end{thebibliography}

\end{document}